\documentclass[prd,preprintnumbers,superscriptaddress,twocolumn,aps,amsmath,amssymb,showpacs,nofootinbib]{revtex4-2}
\pdfoutput=1
\usepackage{graphicx,array}
\usepackage{hyperref}
\usepackage{color}
\usepackage{amsmath,mathrsfs,amssymb,slashed,latexsym}
\usepackage{bbm}
\usepackage{braket}
\usepackage{placeins}
\usepackage{adjustbox}
\usepackage{subcaption}
\usepackage{enumitem}
\usepackage{comment}
\usepackage[compat=1.1.0]{tikz-feynman}
\usepackage{subcaption}
\usepackage{ragged2e}
\allowdisplaybreaks
\usepackage{bm}
\usepackage{siunitx}

\newcommand{\GeV}{\rm GeV}
\definecolor{darkblue}{cmyk}{1,0.4,0,0.3}
\definecolor{violet}{cmyk}{0,1,0,0.2}
\hypersetup{colorlinks, bookmarksnumbered, citecolor=darkblue, linkcolor=darkblue, pdfstartview=FitH, urlcolor=darkblue, linktocpage}

\usepackage{lipsum}
\begin{document}

\title{Geometry-induced azimuthal anisotropy in coherent $J/\psi$ photoproduction}
	
\author{Ding Yu Shao}
\email{dingyu.shao@cern.ch}
\affiliation{Department of Physics, Center for Field Theory and Particle Physics, Key Laboratory of
Nuclear Physics and Ion-beam Application (MOE), Fudan University, Shanghai, 200433, China}
\affiliation{Shanghai Research Center for Theoretical Nuclear Physics, NSFC and Fudan University, Shanghai 200438, China}

\author{Han-Qing Yu}
\email{yuhq24@m.fudan.edu.cn}
\affiliation{Department of Physics, Center for Field Theory and Particle Physics, Key Laboratory of
Nuclear Physics and Ion-beam Application (MOE), Fudan University, Shanghai, 200433, China}

\author{Cheng Zhang}
\email{chengzhang@hznu.edu.cn}
\affiliation{School of Physics, Hangzhou Normal University, Hangzhou, Zhejiang 311121, China}

\author{Jian Zhou}
\email{jzhou@sdu.edu.cn}
\affiliation{Key Laboratory of
Particle Physics and Particle Irradiation (MOE), Institute of
Frontier and Interdisciplinary Science, Shandong University,
(QingDao), Shandong 266237, China \vspace{0.1cm}}
\affiliation{Southern Center for Nuclear-Science Theory (SCNT), Institute of Modern Physics, Chinese Academy of Sciences, HuiZhou, Guangdong
516000, China\vspace{0.1cm}}


\begin{abstract}
Azimuthal anisotropies in heavy-ion collisions are conventionally interpreted as signatures of hydrodynamic flow. We demonstrate that in peripheral collisions, a significant $\cos 2\phi$ asymmetry in the decay leptons of coherently photoproduced $J/\psi$ mesons arises purely from the initial-state geometry of the nuclear electromagnetic field. This modulation originates from the linear polarization of coherent photons, which is radially aligned in impact parameter space and transferred to the vector meson. By employing light-cone perturbation theory within the dipole formalism, we calculate the centrality dependence of this asymmetry for collisions at RHIC and LHC energies. Our predictions quantitatively reproduce STAR data. This observable thus provides a rigorous benchmark for distinguishing electromagnetic initial-state effects from collective medium dynamics. 
\end{abstract}

\maketitle

\section{Introduction}\label{sec:intro}

The azimuthal anisotropy of final-state particles in relativistic heavy-ion collisions is a cornerstone for diagnosing the collective dynamics and transport properties of the Quark-Gluon Plasma (QGP). In central and semi-central collisions, these anisotropies—quantified by Fourier coefficients $v_n$ relative to the reaction plane—are successfully described by the hydrodynamic response of the low-viscosity medium to initial spatial eccentricities~\cite{Heinz:2013th, Romatschke:2017ejr, Gale:2013da, Ollitrault:1992bk, Alver:2010gr, Jia:2024xzx}. However, the observation of similar anisotropic signatures in photon-induced processes, where no bulk medium is expected to form, challenges this strictly hydrodynamic paradigm. This discrepancy necessitates a rigorous discrimination between final-state collective flow and correlations arising purely from the initial-state geometry. Peripheral heavy-ion collisions, where hadronic overlap is minimal yet the electromagnetic fields are intense, provide the ideal laboratory to isolate and study these initial-state geometric effects.

The impact of collision geometry transcends particle yields, deeply influencing spin degrees of freedom, a connection first established in seminal works that linked orbital angular momentum of QGP to final state produced particles' spin polarization~\cite{Liang:2004ph, Liang:2004xn}. This phenomenon is exemplified by the global polarization of $\Lambda$ hyperons and vector mesons~\cite{STAR:2017ckg, STAR:2022fan}, where the system's angular momentum couples to particle spin through the vorticity of the hydrodynamic medium~\cite{Betz:2007kg, Gao:2007bc, Becattini:2007sr}. Just as orbital-spin coupling aligns spin relative to the reaction plane in the bulk medium, the intense electromagnetic (EM) fields in non-central collisions possess a geometric alignment locked to the impact parameter. Thus, distinguishing between vorticity-driven polarization and EM-induced spin phenomena allows one to decouple the properties of the thermalized QGP~\cite{Niida:2024ntm, Skokov:2009qp, Becattini:2021iol, Sheng:2022wsy, Xia:2020tyd, Pang:2016igs, Becattini:2017gcx, Becattini:2021iol, Liu:2019krs, Li:2022pyw, Liang:2019pst, Gao:2012ix, Fu:2021pok, Gao:2021rom, Sheng:2022wsy} from the coherence of the initial state.

Ultraperipheral collisions (UPCs), characterized by impact parameters exceeding twice the nuclear radius, have historically served as the primary domain for studying photon-induced processes driven by the intense electromagnetic fields of relativistic ions~\cite{Baltz:2001dp, STAR:2019wlg, Jia:2024hen, Riffero:2024pmc, STAR:2017enh, Wang:2023yis, CMS:2018bbk, ATLAS:2020epq, ALICE:2019tqa, CMS:2024ykx, Wu:2022exl, Klein:2020jom, Xiao:2020ddm, Xing:2020hwh, Wang:2022gkd, Zhang:2022tee, Zhou:2022twz, Zhao:2022dac, Li:2023qgj, Shuai:2023xvn, Ma:2023dac, Xie:2025sfx}. In this clean environment, hadronic interactions are suppressed, simplifying the theoretical treatment~\cite{Baltz:2007kq}. However, recent measurements by the STAR collaboration in isobaric collisions (Ru+Ru and Zr+Zr) have revealed unexpected and significant $\cos 2\phi$ modulations in coherent $J/\psi$ yields that persist from peripheral into semi-central collisions~\cite{STAR:2017enh, Wang:2023yis}. This observation indicates that coherent photoproduction survives even in the presence of hadronic overlap, providing a window into the nuclear gluon structure~\cite{Klein:1999qj, Guzey:2013jaa} under complex background conditions. Despite these advances, a rigorous theoretical description of the angular correlations in this mixed regime—specifically the coupling between the photoproduced meson and the reaction plane determined by the hadronic bulk—remains to be further developed~\cite{Xiao:2020ddm,Wu:2022exl}.

\begin{figure}[t]
    \centering
    
    \begin{subfigure}{.49\linewidth}
        \centering
        \hspace{-0.5cm}\includegraphics[width=1.1\linewidth]{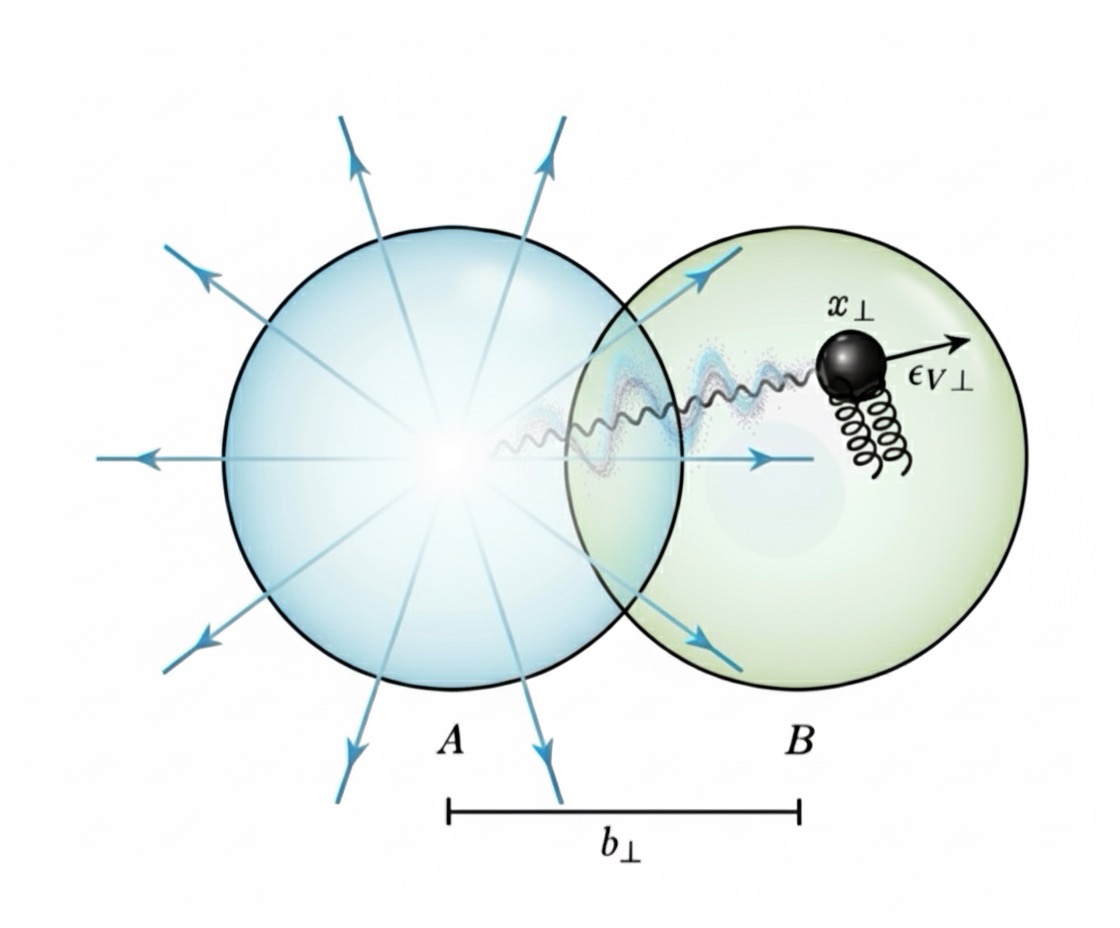}
        \caption{ }
        \label{fig:subfig1}
    \end{subfigure}
    \hfill
    \begin{subfigure}{.49\linewidth}
        \centering
        \includegraphics[width=1.1\linewidth]{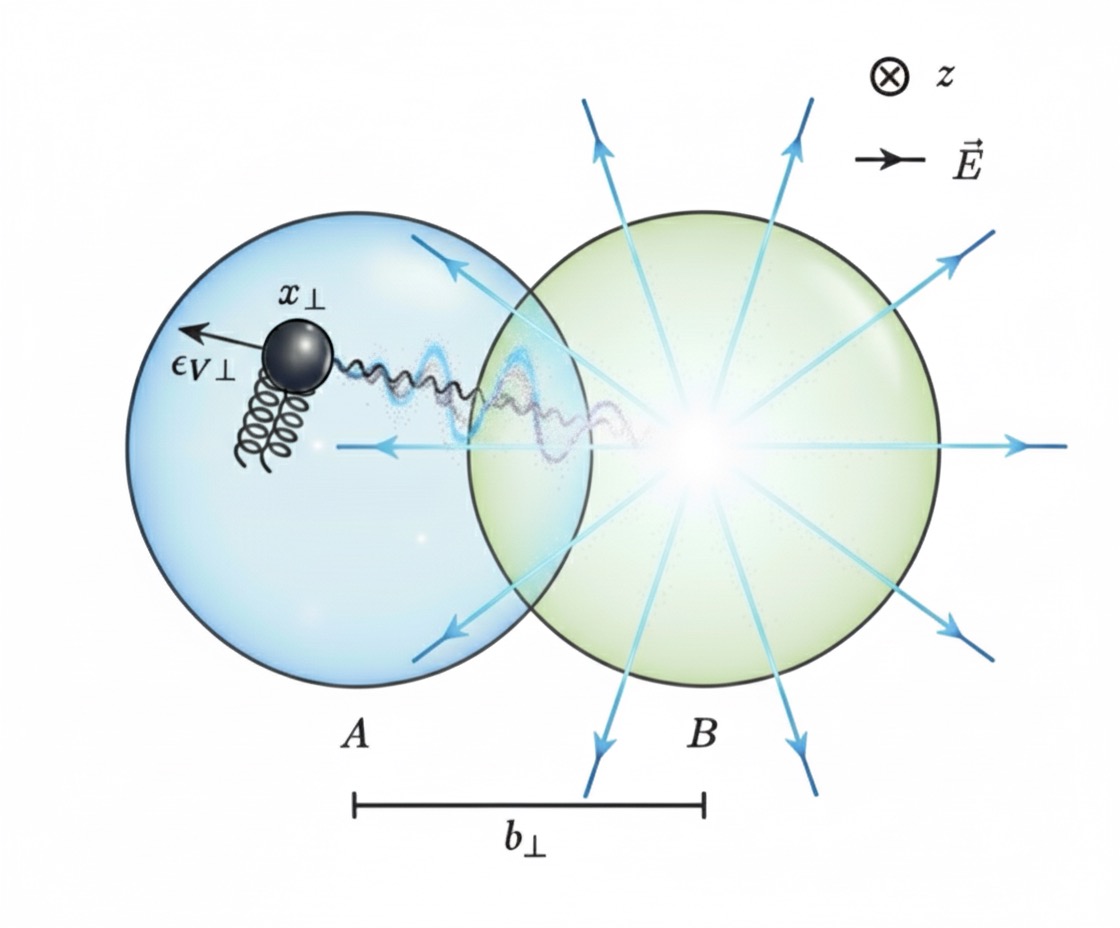}
        \caption{ }
        \label{fig:subfig2}
    \end{subfigure}

    \caption{\justifying Schematic illustration of coherent $J/\psi$ photoproduction in peripheral heavy ion collisions. In both panels, $b_{\perp}$ denotes the impact parameter, $x_{\perp}$ marks the transverse production point, and $\epsilon_{V\perp}$ indicates the transverse polarization vector of the $J/\psi$.}
    \label{fig:combined}
\end{figure}

In this paper, we investigate the azimuthal anisotropy of leptons arising from the decay of coherently photoproduced $J/\psi$ mesons in peripheral heavy-ion collisions, as depicted in Fig.~\ref{fig:combined}. The underlying mechanism is the transfer of polarization from the initial state to the final particles: coherent photons carry linear polarization radially aligned with the impact parameter vector. This polarization is inherited by the produced $J/\psi$ and subsequently analyzed via its leptonic decay~\cite{Li:2019sin, Klein:2020jom}. A key advantage of peripheral collisions over pure UPCs is the ability to experimentally reconstruct the reaction plane using spectator nucleons or bulk elliptic flow. This defines a reference axis for the lepton transverse momentum $\boldsymbol{P}_\perp$, resulting in a predictable $\cos 2\phi$ modulation. While conceptually similar to the $\cos 4\phi$ asymmetry in $\gamma \gamma \to \ell^+ \ell^-$~\cite{Xiao:2020ddm}, the intermediate spin-1 vector meson dictates a $\cos 2\phi$ signature, providing a unique probe of the photon polarization vector.

We compute the coherent $J/\psi$ photoproduction amplitude within the framework of light-cone perturbation theory (LCPT)~\cite{Langnau:1992gk, Brodsky:1997de, Srivastava:2000zz, Mantovani:2016uxq, Hanninen:2017ddy, Beuf:2021srj}, employing the dipole formalism and the IP-Sat saturation model to describe the nuclear gluon distribution. By mapping the impact parameter to centrality via an optical Glauber model, we determine the centrality dependence of the $\cos 2\phi$ asymmetry. Our calculations reproduce the rising asymmetry observed at RHIC and provide predictions for LHC energies. These findings identify the polarization of the initial photon flux as a robust source of azimuthal anisotropy, distinguishable from hydrodynamic flow, thus offering a vital baseline for disentangling initial-state geometry from medium properties.

The paper is organized as follows. Sec.~\ref{sec:theoretical} details the LCPT framework for calculating production amplitudes and anisotropies. Sec.~\ref{sec:numerical} describes the numerical implementation, including the dipole and centrality models. Results for RHIC and LHC energies are presented in Sec.~\ref{sec:results}, followed by conclusions in Sec.~\ref{sec:conc}.

\section{Theoretical Framework}\label{sec:theoretical}

In this section, we describe the theoretical framework for calculating the spin asymmetry in peripheral collisions, focusing on the vector meson production process $A+A \rightarrow V +A^\prime+A^\prime$. In this process, one nucleus serves as a source of photons that scatter coherently off the other nucleus~\cite{Zhou:2024acp, Hu:2024bsm, Schenke:2024gnj, Zhang:2024mql, Baier:2004tj, Klein:1999qj, Goncalves:2005sn}. The photon-nuclear interaction is modeled using the dipole picture: the incoming photon first splits into a quark-antiquark pair, which then elastically scatters off the color glass condensate (CGC) gluons in the target nucleus~\cite{Dumitru:2002qt, Hatta:2016dxp}, before recombining to form a vector meson~\cite{Gelis:2002fw, Gelis:2002ki}.

The amplitude for the gluon transition can be written as
\begin{align}\label{eq:gluon_dis}
    \Tilde{\mathcal{T}}(\boldsymbol{x}_\perp,\xi_g)=i\int&\frac{d^2\boldsymbol{r}_\perp}{4\pi}\int^1_0 dz\, \Psi_{\gamma\rightarrow q\bar{q}}(\boldsymbol{r}_\perp,z,\boldsymbol{\epsilon}_{\gamma\perp})\notag\\&\times N(\boldsymbol{r}_\perp,\boldsymbol{x}_\perp)\Psi^*_{V\rightarrow q\bar{q}}(\boldsymbol{r}_\perp,z,\boldsymbol{\epsilon}_{V\perp}),
\end{align}
where $N(\boldsymbol{r}_\perp, \boldsymbol{x}_\perp)$ is the dipole amplitude for a $q\bar{q}$ dipole of size $\boldsymbol{r}_\perp$ scattering off the target nucleus at impact parameter $\boldsymbol{x}_\perp$ in the photon-nuclear collision. $\xi_g$ is the momentum fraction of the gluon, and its dependence is included in $N(\boldsymbol{r}_\perp, \boldsymbol{x}_\perp)$. 

The polarization-dependent wave functions $\Psi_{\gamma \to q\bar{q}}$ and $\Psi_{V \to q\bar{q}}$ for the photon and vector meson, respectively, are determined from LCPT~\cite{Kowalski:2003hm,Kowalski:2006hc}
\begin{align}
\Psi_{\gamma\rightarrow q\bar{q}}(\boldsymbol{r}_\perp,&z,\boldsymbol{\epsilon}_{\gamma\perp})=\frac{e e_q}{2\pi} \delta_{aa^{\prime}} \big\{ \delta_{\sigma,-\sigma^{\prime}} [(1-2z)i\boldsymbol{\epsilon}_{\gamma\perp}\cdot\boldsymbol{r}_\perp\notag
\\&+\sigma\boldsymbol{\epsilon}_{\gamma\perp}\times\boldsymbol{r}_\perp]\frac{-1}{|\boldsymbol{r}_\perp|}\frac{\partial}{\partial|\boldsymbol{r}_\perp|}\notag
\\&+\delta_{\sigma\sigma^{\prime}}m_q(\sigma\boldsymbol{\epsilon}_{\gamma\perp}^{(1)}+i\sigma\boldsymbol{\epsilon}_{\gamma\perp}^{(2)})\big\}K_0(|\boldsymbol{r}_\perp|e_f),
\end{align}
\begin{align}
\Psi_{V\rightarrow q\bar{q}}(\boldsymbol{r}_\perp,&z,\boldsymbol{\epsilon}_{V\perp})=\delta_{aa^{\prime}} \big\{ \delta_{\sigma,-\sigma^{\prime}} [(1-2z)i\boldsymbol{\epsilon}_{V\perp}\cdot\boldsymbol{r}_\perp\notag
\\&+\sigma\boldsymbol{\epsilon}_{V\perp}\times\boldsymbol{r}_\perp]\frac{-1}{|\boldsymbol{r}_\perp|}\frac{\partial}{\partial|\boldsymbol{r}_\perp|}\notag
\\&+\delta_{\sigma\sigma^{\prime}}m_q(\sigma\boldsymbol{\epsilon}_{V\perp}^{(1)}+i\sigma\boldsymbol{\epsilon}_{V\perp}^{(2)})\big\}\Phi(|\boldsymbol{r}_\perp|,z).
\end{align}
Here, $\sigma$ ($\sigma^\prime$) denotes the helicities of the quark and antiquark, and $a$ ($a^\prime$) are color indices. $m_q$ and $e_q$ are the mass and electric charge of the quark with flavor $q$, respectively. The function $K_0$ in $\Psi_{\gamma\rightarrow q\bar{q}}$ is the modified Bessel function of the second kind, and $e_f$ is defined as $e_f^2 \equiv Q^2 z (1-z) + m_q^2$ with $Q^2 = \boldsymbol{k}_\perp^2 + \xi^2 M_p^2$, where $M_p$ is the proton mass and $\xi$ is the momentum fraction. $\Phi(r_\perp, z)$ is the scalar part of the vector meson wave function, which will be specified shortly. $z$ is the longitudinal momentum fraction carried by the quark, and $\boldsymbol{\epsilon}_{\gamma\perp}$ ($\boldsymbol{\epsilon}_{V\perp}$) are the transverse polarization vectors of the photon (vector meson).

We observe that the amplitude in Eq.(\ref{eq:gluon_dis}) carries an explicit dependence on the transverse polarization vectors through the light-cone wave functions. It is therefore convenient to factor out the universal polarization structure and define a scalar amplitude $\Tilde{\mathcal{A}}(\boldsymbol{x}_\perp,\xi_g)$ that contains only the part requiring numerical evaluation
\begin{align}
    \Tilde{\mathcal{A}}(\boldsymbol{x}_\perp,\xi_g)&=i\frac{N_c e e_q}{\pi}\int \frac{d^2\boldsymbol{r}_\perp}{4\pi}N(\boldsymbol{r}_\perp,\boldsymbol{x}_\perp)\notag\\
    &\times\int^1_0 dz\bigg\{m_q^2\Phi^*(|\boldsymbol{r}_\perp|,z)K_0(|\boldsymbol{r}_\perp|e_f)\\
    &+[z^2+(1-z)^2]\frac{\partial\Phi^*(|\boldsymbol{r}_\perp|,z)}{\partial|\boldsymbol{r}_\perp|}\frac{\partial K_0(|\boldsymbol{r}_\perp|e_f)}{\partial|\boldsymbol{r}_\perp|} \bigg\},\notag
\end{align}
which represents the overlap between the photon and vector meson wave functions modulated by the dipole amplitude, and we have $\Tilde{\mathcal{T}}(\boldsymbol{x}_\perp,\xi_g)=\Tilde{\mathcal{A}}(\boldsymbol{x}_\perp,\xi_g)\boldsymbol{\epsilon}_{\gamma\perp}\cdot\boldsymbol{\epsilon}_{V\perp}$.

In a $A+A$ collision, the photon source is ambiguous, leading to interference between emission from either nucleus. The total amplitude for vector meson production at transverse position $\boldsymbol{x}_\perp$ and internuclear impact parameter $\boldsymbol{b}_\perp$ is the coherent sum of these two contributions
\begin{align}
    &\Tilde{\mathcal{M}}_{\gamma A\rightarrow V}(\boldsymbol{x}_\perp,\,\boldsymbol{b}_\perp)=\Tilde{\mathcal{A}}(\boldsymbol{x}_{\perp}-\boldsymbol{b}_\perp,\xi_1)\Tilde{\mathcal{F}}(\boldsymbol{x}_{\perp},\xi_2)\,\hat{\boldsymbol{x}}_{\perp}\cdot\boldsymbol{\epsilon}_{V\perp}\notag\\
    &\qquad+\Tilde{\mathcal{F}}(\boldsymbol{x}_{\perp}-\boldsymbol{b}_\perp,\xi_1)\Tilde{\mathcal{A}}(\boldsymbol{x}_{\perp},\xi_2)\,\frac{(\boldsymbol{x}_{\perp}-\boldsymbol{b}_\perp)\cdot\boldsymbol{\epsilon}_{V\perp}}{|\boldsymbol{x}_{\perp}-\boldsymbol{b}_\perp|}.
\end{align}
Here, we use the fact that in the coordinate-space representation, the quasi-real photon emitted by a relativistic nucleus is characterized by a transverse electromagnetic field that is purely radial, and the effective linear-polarization direction of the photon in the small-$x$ limit is taken to be along the radial unit vector, that is $\boldsymbol{\epsilon}_{\gamma\perp} = \hat{\boldsymbol{x}}_\perp = \boldsymbol{x}_\perp / |\boldsymbol{x}_\perp|$. Moreover, $\Tilde{\mathcal{F}}$ represents the photon distribution function, the variable $\xi$ denotes the longitudinal momentum fraction transferred to the meson, and $\boldsymbol{b}_\perp$ is the impact parameter between the two colliding nuclei.

The momentum space representation of the production amplitude is given by
\begin{align}\label{eq:amplitude}
    &\mathcal{M}_{\gamma A\rightarrow V}(\boldsymbol{q}_\perp,\boldsymbol{b}_\perp)=\int d^2 \boldsymbol{x}_\perp e^{i \boldsymbol{q}_\perp\cdot\boldsymbol{x}_\perp}\Tilde{\mathcal{M}}_{\gamma A\rightarrow V}(\boldsymbol{x}_\perp,\boldsymbol{b}_\perp)\notag\\
    &=\int d^2 \boldsymbol{x}_\perp\big[e^{i\boldsymbol{q}_\perp\cdot\boldsymbol{x}_\perp}\Tilde{\mathcal{F}}(\boldsymbol{x}_\perp,\xi_2)\Tilde{\mathcal{A}}(\boldsymbol{x}_\perp-\boldsymbol{b}_\perp,\xi_1)\\
    &\quad-e^{i\boldsymbol{q}_\perp\cdot(\boldsymbol{b}_\perp-\boldsymbol{x}_\perp)}\Tilde{\mathcal{F}}(-\boldsymbol{x}_\perp,\xi_1)\Tilde{\mathcal{A}}(\boldsymbol{b}_\perp-\boldsymbol{x}_\perp,\xi_2)\big]\hat{\boldsymbol{x}}_\perp\cdot\boldsymbol{\epsilon}^*_{V\perp}.\notag
\end{align}
where $\boldsymbol{q}_\perp$ is the transverse momentum of the produced vector meson. 

We focus on the leptonic decay channel $J/\psi \to \ell^+ \ell^-$. The polarization of the vector meson is analyzed via the momenta of the decay leptons, $\boldsymbol{p}_{1\perp}$ and $\boldsymbol{p}_{2\perp}$. We have the pair transverse momentum $\boldsymbol{q}_\perp = \boldsymbol{p}_{1\perp} + \boldsymbol{p}_{2\perp}$ and the relative momentum $\boldsymbol{P}_\perp = (\boldsymbol{p}_{1\perp} - \boldsymbol{p}_{2\perp})/2$. The decay amplitude can be described by
\begin{align}
     &\mathcal{M}_{\gamma A\rightarrow J/\psi\rightarrow\ell^+\ell^-}\notag\\
     &=\frac{\epsilon^\mu_\perp \bar{u}(p_1)\gamma_\mu v(p_2)}{M^2_{\ell^+\ell^-}-M^2_{J/\psi}+i M_{J/\psi}\Gamma_{J/\psi}}\mathcal{M}_{\gamma A\rightarrow J/\psi},
\end{align}
where $\epsilon_\perp^\mu = (0, \boldsymbol{\epsilon}_{V\perp}, 0)$. $M_{\ell^+\ell^-}$ is the invariant mass of the lepton pair. $M_{J/\psi}$ and $\Gamma_{J/\psi}$ are the mass and decay width of the $J/\psi$ particle, respectively. The factor $\hat{\boldsymbol{x}}_\perp \cdot \boldsymbol{\epsilon}^*_{V\perp}$ in the production amplitude \eqref{eq:amplitude} translates, after the spin sum, into a correlation $\hat{\boldsymbol{x}}_\perp \cdot \boldsymbol{P}_\perp$ in the lepton cross section. We then obtain
\begin{align}
    &\frac{d\sigma_{\gamma A\rightarrow J/\psi \rightarrow\ell^+\ell^-}}{d^2\boldsymbol{q}_\perp d^2\boldsymbol{P}_\perp dy_1 dy_2 d^2\boldsymbol{b}_\perp}=|\mathcal{M}_{\gamma A\rightarrow J/\psi\rightarrow\ell^+\ell^-}|^2\notag\\
    &=\int d^2\boldsymbol{x}_\perp d^2\boldsymbol{x}_\perp^\prime\hspace{1pt}\mathcal{K}(\boldsymbol{x}_\perp,\boldsymbol{x}_\perp^\prime,\boldsymbol{b}_\perp,\boldsymbol{q}_\perp)\notag\\
    &\hspace{10pt}\times\left[M^2_{J/\psi}\hat{\boldsymbol{x}}_{\perp}\cdot\hat{\boldsymbol{x}}_{\perp}^\prime-4(\hat{\boldsymbol{x}}_{\perp}\cdot\boldsymbol{P}_\perp)(\hat{\boldsymbol{x}}_{\perp}^\prime\cdot\boldsymbol{P}_\perp)\right],
    \label{eq:crosssection}
\end{align}
where $y_1$ and $y_2$ are the rapidities of the final state leptons, and $\mathcal{K}$ denotes the kernel function that contains the distribution functions, phase factors, and other scalar components of the cross section, which can be derived from Eq.~\eqref{eq:amplitude}. The term $(\hat{\boldsymbol{x}}_\perp \cdot \boldsymbol{P}_\perp)(\hat{\boldsymbol{x}}_\perp^\prime \cdot \boldsymbol{P}_\perp)$ generates the azimuthal modulation. Specifically, we isolate the $\cos 2\phi$ component using the decomposition as

\begin{align}
    &(\hat{\boldsymbol{x}}_\perp\cdot \hat{\boldsymbol{P}}_\perp)(\hat{\boldsymbol{x}}^\prime_\perp\cdot \hat{\boldsymbol{P}}_\perp)=\frac{1}{2}\Big[ \cos(\phi_{\boldsymbol{x}_\perp}-\phi_{\boldsymbol{x}^\prime_\perp})\notag\\
&\quad+\cos(\phi_{\boldsymbol{x}_\perp}+\phi_{\boldsymbol{x}^\prime_\perp}-2\phi_{\boldsymbol{b}_\perp})\cos(2\phi)\notag\\
&\quad-\sin{(\phi_{\boldsymbol{x}_\perp}+\phi_{\boldsymbol{x}^\prime_\perp}-2\phi_{\boldsymbol{b}_\perp})}\sin{(2\phi)}\Big],
\end{align}
where we define $\phi \equiv \phi_{\boldsymbol{b}_\perp} - \phi_{\boldsymbol{P}_\perp}$. It is noted that the sine terms vanish upon integration over $\boldsymbol{x}_\perp$.    

\section{Numerical implementation}\label{sec:numerical}
In this section, we discuss all ingredients necessary for numerical estimation of the $\cos 2\phi$ asymmetry for photoproduction of $J/\psi$ in peripheral heavy ion collisions. We begin with introducing the model input for the dipole amplitude $N(\boldsymbol{r}_\perp, \boldsymbol{x}_\perp)$
\begin{align}    &N(\boldsymbol{r}_\perp,\boldsymbol{x}_\perp)=\notag\\
    &\qquad1-\frac{1}{N_c}\left\langle Tr\left(U(x_\perp+\frac{r_\perp}{2})U^\dagger(x_\perp-\frac{r_\perp}{2})\right)\right\rangle.
\end{align}
In the small $x$ limit, the expectation value of the dipole amplitude can be calculated using a semi-classical model~\cite{McLerran:1993ka, McLerran:1993ni}. Here, we instead use a phenomenological expression of the amplitude~\cite{Kowalski:2003hm, Kowalski:2006hc} for purpose of a more practical phenomenological study
\begin{align}
    N(\boldsymbol{r}_\perp,\boldsymbol{x}_\perp)=1-e^{-2\pi B_p A T_A(\boldsymbol{x}_\perp) \mathcal{N}(\boldsymbol{r}_\perp)},
\end{align}
where $A$ is the nuclear atomic number and $B_p=4\ \GeV^{-2}$ in IP-Sat model. $T_A(\boldsymbol{x}_\perp)$ is the nuclear thickness function, which gives the distribution of the nucleons inside a nucleus in the transverse plane, which is determined using the Woods-Saxon distribution in numerical calculation. Besides, we also employ a uniform-disc parametrization in part of our numerical analysis. In this simplified model, $T_A(\boldsymbol{x}_\perp)$ is taken to be a constant for $|\boldsymbol{x}_\perp|<R_A$ and zero otherwise. This auxiliary parametrization is introduced to obtain a controlled lower bound for the resulting $\cos 2\phi$ asymmetry. $\mathcal{N}(\boldsymbol{r}_\perp)$ is the dipole-nucleon scattering amplitude, and in a modified IP-Sat model~\cite{Lappi:2010dd, Lappi:2013am} it can be written as
\begin{align}
    \mathcal{N}(\boldsymbol{r}_\perp)=1-\exp{\left[-\boldsymbol{r}_{\perp}^2 G(\xi_g,\boldsymbol{r}_\perp)\right]}.
\end{align}
Here $G(\xi_g,\boldsymbol{r}_\perp)$ is proportional to the DGLAP evolved gluon distribution in the Bartels, Golec, Biernat, and Kowalski (BGBK) parametrization. To simplify calculation, we adopt another parametrization known as the Golec-Biernat and Wusthoff (GBW) model~\cite{Golec-Biernat:1998zce, Golec-Biernat:1999qor}
\begin{align}
    G(\xi_g)=\frac{1}{4}Q_s^2(\xi_g),
\end{align}
where $Q_s(\xi_g)=(\xi_0/\xi_g)^{\lambda_{\rm GBW}/2}$ is the saturation scale. The parameters we use are $\xi_0=3\times10^{-4}$ and $\lambda_{\rm GBW}=0.29$, which were determined by fitting to HERA data~\cite{Kowalski:2006hc}.

The photon distribution amplitude $\mathcal{F}(\boldsymbol{k}_\perp, \xi)$ can be derived within the equivalent photon approximation~\cite{PhysRevC.47.2308, Bertulani:1987tz} 
\begin{align}
    \mathcal{F}(\boldsymbol{k}_\perp,\xi)=\frac{Z\sqrt{\alpha_e}}{\pi}|\boldsymbol{k}_\perp|\frac{F(\boldsymbol{k}^2_\perp+\xi^2M^2_p)}{\boldsymbol{k}^2_\perp+\xi^2M^2_p},
\end{align}
where $M_p$ is the proton mass. The nuclear charge form factor $F$ is taken from the STARlight Monte Carlo generator~\cite{Klein:2016yzr}
\begin{align}
    &F(|\boldsymbol{k}|^2)=\\
    &\frac{3}{|\boldsymbol{k}|^3 R^3_A}\left[\sin{(|\boldsymbol{k}|R_A)}-|\boldsymbol{k}|R_A \cos{(|\boldsymbol{k}|R_A)}\right]\frac{1}{a^2|\boldsymbol{k}|^2+1},\notag
\end{align}
with $R_A = 1.1 A^{1/3}$ fm and $a = 0.7$ fm. We compute $|\boldsymbol{k}| = \sqrt{\boldsymbol{k}^2_\perp + \xi^2 M_p^2}$ to evaluate the photon distribution. Although this distribution is initially in momentum space, we perform a Fourier transformation to obtain the position-dependent distribution $\Tilde{\mathcal{F}}(\boldsymbol{x}_\perp, \xi)$.

For the scalar part of vector meson wave function, we use the ``Gaus-LC" wave function~\cite{Kowalski:2003hm, Kowalski:2006hc}
\begin{align}
    \Phi(|\boldsymbol{r}_\perp|,z)=\beta z(1-z)\exp{\Bigg[-\frac{\boldsymbol{r}_\perp^2}{2R_\perp^2}\Bigg]},
\end{align}
with parameters $\beta = 1.23$ and $R_\perp^2 = 6.5 \, \GeV^{-2}$ for the $J/\psi$.

As the impact parameter $\boldsymbol{b}_\perp$—the transverse distance between the centers of the two colliding nuclei—cannot be directly measured, centrality is commonly used to characterize the geometry of collisions in experiments. Centrality quantifies the degree of overlap between the nuclei and can be inferred from observables such as particle multiplicity or transverse energy. In this work, we employ the optical Glauber model~\cite{Glauber:1955qq, Miller:2007ri, PHOBOS:2007vdf, Loizides:2014vua, Broniowski:2007nz} to relate the collision centrality to the impact parameter. This semi-classical model describes the probabilistic distribution of nucleons in the nuclei and their interactions, linking the geometric parameter $b_\perp$ to experimental observables. In this model, the nucleons in each nucleus are assumed to follow a Woods-Saxon distribution~\cite{DeVries:1987atn}
\begin{align}
    \rho_A(r)=\frac{C_0}{1+\exp\left[(r-R_{WS})/a\right]},
\end{align}
where $R_{WS} = 6.38$ fm and $a = 0.535$ fm for Au nuclei, and $R_{WS} = 6.62$ fm and $a = 0.546$ fm for Pb nuclei~\cite{Miller:2007ri, PHENIX:2004vdg}. The normalization constant $C_0$ ensures that the distribution integrates to the mass number $A$.

\begin{figure}[t]
    \centering
    \includegraphics[width=1.0\linewidth]{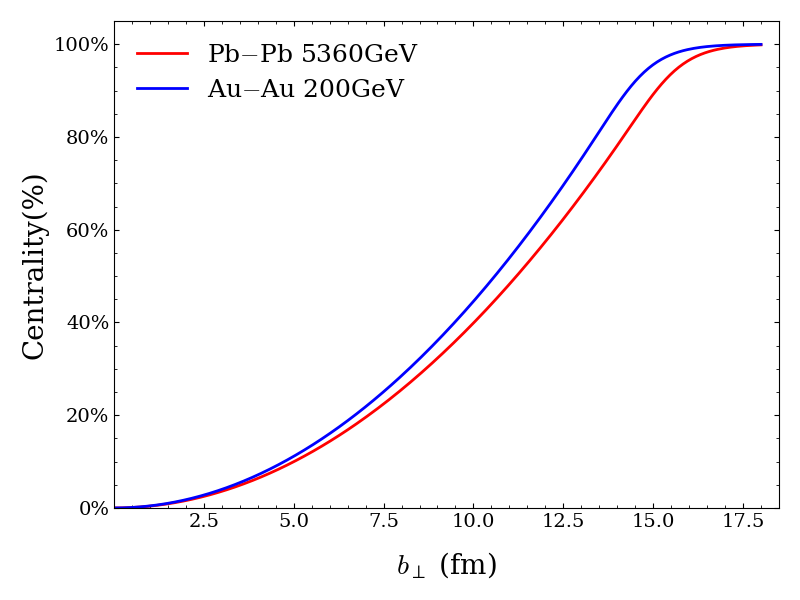}
    \caption{\justifying The centrality $c(b_\perp)$  computed using the optical Glauber model is plotted as a function of impact parameter $b_\perp$.}
        \label{fig:centrality}
\end{figure}

The nuclear thickness function $T_A(\boldsymbol{s})$ is then given by
\begin{align}
    T_A(\boldsymbol{x}_\perp)=\int^\infty_{-\infty} \rho_A(\sqrt{|\boldsymbol{x}_\perp|^2+z^2})dz,
\end{align}
with $C_0$ determined by the condition $\int T_A(\boldsymbol{x}_\perp) \, d^2 \boldsymbol{x}_\perp = A$. Given the thickness function, the overlap function for nucleus-nucleus (A-A) collisions is
\begin{align}
    T_{AA}(\boldsymbol{b}_\perp)=\int T_A(\boldsymbol{x}_\perp)T_A(\boldsymbol{x}_\perp+\boldsymbol{b}_\perp)d^2\boldsymbol{x}_\perp.
\end{align}
Employing the optical approximation, the probability of having at least one inelastic nucleon-nucleon collision at $\boldsymbol{b}_\perp$ is given by
\begin{align}
    P_{in}(\boldsymbol{b}_\perp)=1-\exp\left[-\sigma_{NN}T_{AA}(\boldsymbol{b}_\perp)\right],
\end{align}
where $\sigma_{NN}$ is the inelastic nucleon-nucleon cross section. According to the recent studies~\cite{STAR:2020phn, CMS:2024ykx}, the best fit for $\sigma_{NN}$  is $42 \pm 3$ mb and $68 \pm 1.2$ fm for the center of mass energy 200 GeV and 5360 GeV, respectively.

The differential inelastic cross section is then expressed as,
\begin{align}
    \frac{d\sigma_{in}}{d|\boldsymbol{b}_\perp|}=2\pi| \boldsymbol{b}_\perp| P_{in}(\boldsymbol{b}_\perp),
\end{align}
which leads to  the definition  of  centrality,
\begin{align}
    c(\boldsymbol{b}_\perp)&=\frac{1}{\sigma_{in}}\int_0^{|\boldsymbol{b}_\perp|}\frac{d\sigma_{in}}{db^\prime}db^\prime \notag \\
    &=\frac{\int^{|\boldsymbol{b}_\perp|}_0 2\pi b^\prime P_{in}(b^\prime)db^\prime}{\int^\infty_0 2\pi b^\prime P_{in}(b^\prime)db^\prime}.
\end{align}
Using this result, we plot the centrality as a function of $b_\perp$ in the Fig.~\ref{fig:centrality}. With all these components introduced above, we are ready to compute the azimuthal dependent cross section. 

\section{Phenomenological results}\label{sec:results}

\begin{figure}[t]
    \centering
    \includegraphics[width=1.\linewidth]{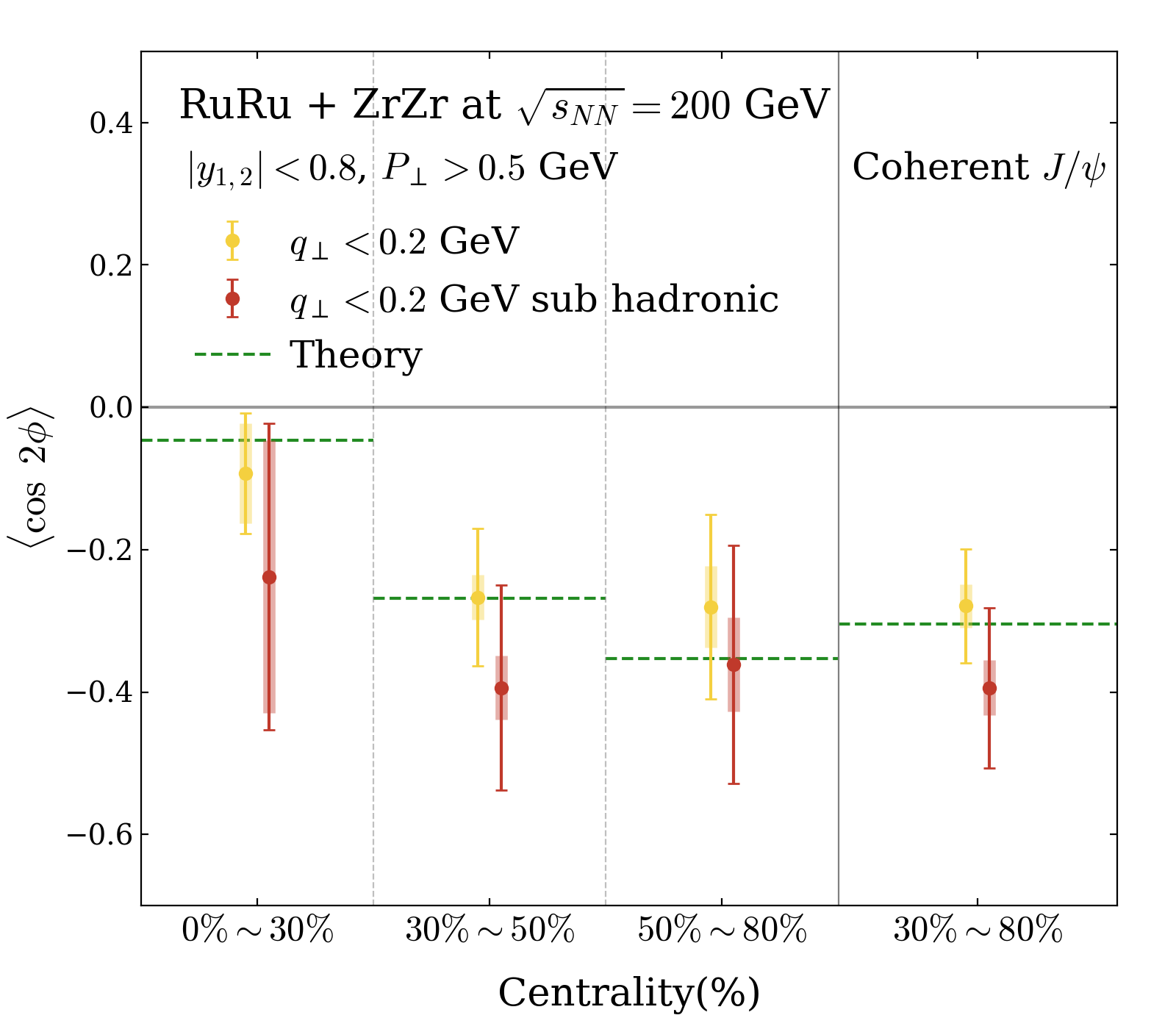}
    \caption{\justifying Centrality dependence of the $\cos 2\phi$ asymmetry in coherent $J/\psi$ production at 200 GeV, compared with STAR measurements~\cite{STAR:2017enh,Wang:2023yis}.  $J/\psi$ is reconstructed via its decay into an $e^+ e^-$ pair.}
    \label{fig:STAR_results}
\end{figure}

In this section, we present the numerical results for coherent $J/\psi$ photoproduction in peripheral heavy-ion collisions, focusing on the centrality dependence of the $\cos 2\phi$ asymmetry.  To quantify the magnitude of the azimuthal asymmetry, we introduce the variable $\cos 2\phi$  by
\begin{align}
    \langle\cos (2\phi)\rangle\equiv\frac{\int d\sigma\cos (2\phi) }{\int d\sigma},
\end{align}
where $d\sigma$ is the differential cross section introduced in Eq.~\eqref{eq:crosssection}, and is integrated over the same phase space region in the numerator and the denominator. We consider Ru+Ru ($Z=44$, $A=96$) and Zr+Zr ($Z=40$, $A=96$) collisions at $\sqrt{s_{NN}} = 200$ GeV (RHIC), and Pb+Pb collisions at $\sqrt{s_{NN}} = 5.36$ TeV (LHC). 

Fig.~\ref{fig:STAR_results} illustrates the centrality dependence of the $\cos 2\phi$ asymmetry at RHIC energy. Our theoretical calculation is consistent with the experimental data from STAR within uncertainties, capturing the rise in asymmetry from central to peripheral collisions.   The asymmetry reaches nearly $40\%$ in the centrality category $[50\%-80\%]$. This increase corresponds to larger impact parameters, where the photon polarization is more aligned with the direction of impact parameter, enhancing the polarization-induced effects. 

\begin{figure}[t]
    \centering
    \includegraphics[width=1.02\linewidth]{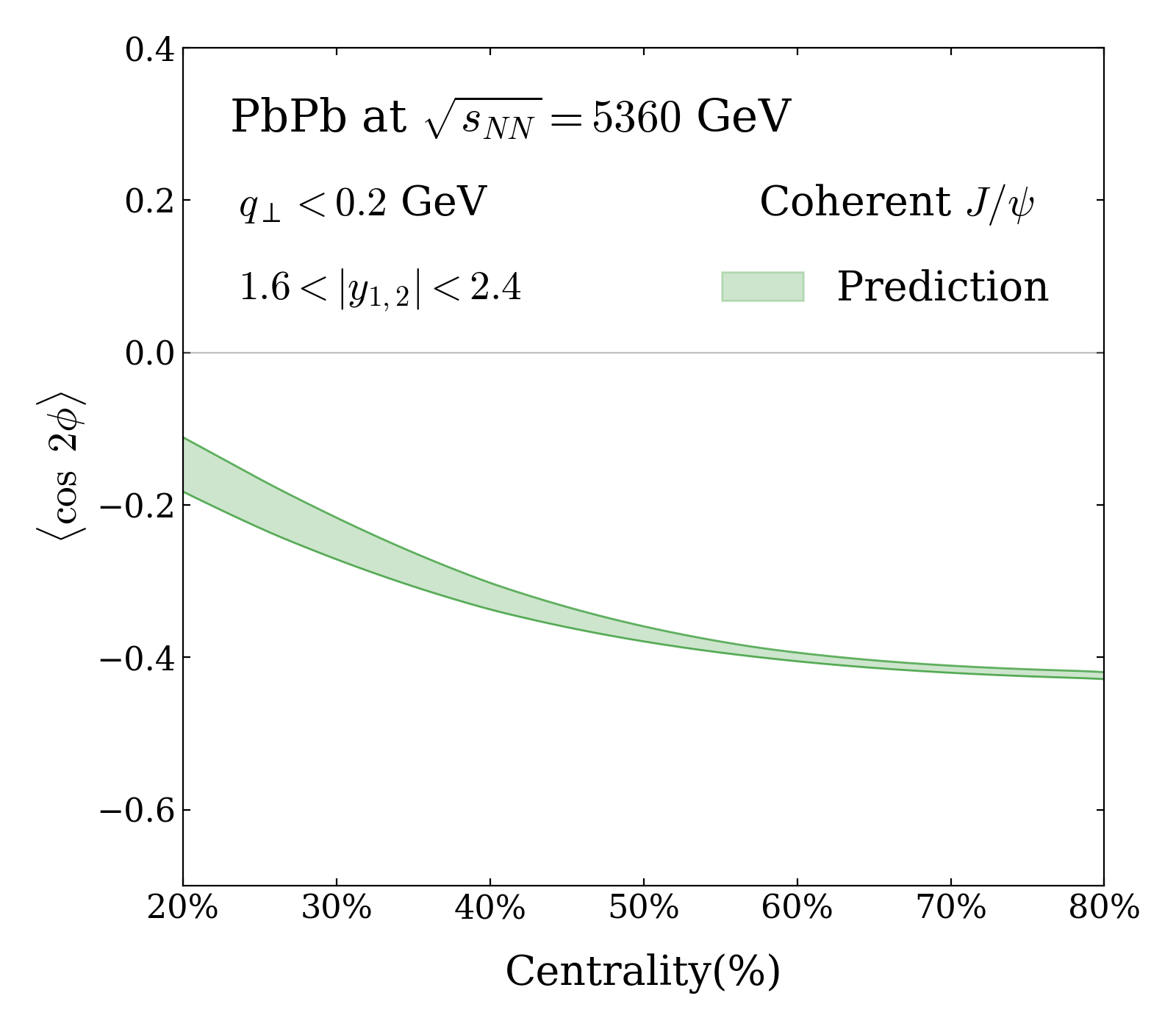}
    \caption{\justifying Predicted centrality dependence of the $\cos 2\phi$ asymmetry in coherent $J/\psi$ production at 5.36 TeV. $J/\psi$ is reconstructed via its decay into a $\mu^+ \mu^-$ pair. The upper, less-negative edge of the band corresponds to the uniform-disc geometry, whereas the lower, more-negative edge is given by the GBW dipole model.}
    \label{fig:CMS_results}
\end{figure}

Extending to LHC energy, Fig.~\ref{fig:CMS_results} shows the predicted $\cos 2\phi$ asymmetry for Pb+Pb collisions, relevant to CMS kinematics~\cite{CMS:2024ykx}. The modulation exhibits a similar centrality dependence to RHIC, with the asymmetry rising gradually and saturating at higher centralities. Although no data are currently available, these predictions can be tested in future peripheral Pb+Pb runs.   The resulting modulation reflects interference between different photon helicity states, and is also sensitive to transverse spatial gluon distribution. 
Despite the similarity between $v_2$ low and the   $\cos 2\phi$ asymmetries under consideration, the underlying mechanism is drastically different. The $\cos2\phi$ asymmetry we considered arises from initial state effect, while $v_2$  flow is a pure final state effect.

\section{Conclusion}\label{sec:conc}
In this work, we have performed a comprehensive analysis of the reaction plane geometry-induced $\cos 2\phi$ azimuthal asymmetry in coherent $J/\psi$ photoproduction within peripheral heavy-ion collisions. By integrating LCPT with the CGC dipole formalism and the IP-Sat saturation model, our calculations successfully reproduce the characteristic features of the asymmetry observed by the STAR Collaboration in Ru+Ru and Zr+Zr collisions at RHIC energies.

We demonstrated that this observed modulation arises naturally from the linear polarization of coherent photons, which are polarized radially in impact parameter space. This initial polarization leads to an interference between photon helicity states, which translates into a significant azimuthal anisotropy in the momenta of the decay leptons relative to the reaction plane. Consequently, our results provide a compelling explanation for the experimental data based purely on electromagnetic and geometric origins, without the need to invoke hydrodynamic flow. Building on these results, we extended our framework to the TeV scale, presenting predictions for Pb+Pb collisions at the LHC. We observe a distinct centrality dependence where the asymmetry increases significantly as the collision system evolves from semi-central to peripheral, eventually saturating. Although experimental data for this specific observable are not yet available at the LHC, these predictions serve as a robust benchmark for future measurements. They are essential for distinguishing the geometric signatures of photonuclear interactions from hadronic flow phenomena, ensuring that observed correlations in peripheral bins are correctly attributed to their physical source.

Overall, this study establishes the $\cos 2\phi$ asymmetry as a precise probe of the transverse structure of the nuclear photon field and small-$x$ gluon dynamics. Our framework provides a transparent theoretical link between the macroscopic collision geometry and the microscopic partonic interactions, offering a method to constrain initial-state anisotropies that is largely free from QGP interference. Experimental confirmation of these predictions would crucially refine our ability to disentangle initial-state geometry from final-state medium responses, a necessary step toward isolating the electromagnetic properties of the quark-gluon plasma. Looking ahead, future developments of this framework should incorporate higher-order perturbative corrections and incoherent diffraction processes to further improve precision. Furthermore, a systematic treatment of final-state interactions with the medium could be included to address potential modifications in more central collisions. Such refinements will be instrumental in fully exploiting the precision of forthcoming data from RHIC and the LHC, thereby advancing our understanding of photon-nucleus interactions in the high-energy limit.

\vspace{3mm}
\noindent{\it Acknowledgments.}
The authors thank Shuai Yang and Wangmei Zha for the helpful discussions. This work has been supported by the National Natural Science Foundations of China under Grant No.~12175118 (J.Z.), No.~12321005 (J.Z.), No.~12275052 (H.Q.Y. and D.Y.S.), No.~12147101 (D.Y.S.) and No.~12405151 (C.Z.). D.Y.S. is also supported by the Innovation Program for Quantum Science and Technology under grant No. 2024ZD0300101.

\bibliographystyle{apsrev4-1}

\bibliography{reference}

\end{document}